\documentclass[aps,prl,floatfix,superscriptaddress,showpacs,twocolumn]{revtex4}
\usepackage[dvips]{graphicx}

\usepackage{longtable}
\usepackage{dcolumn}
\usepackage[dvips]{graphicx}
\usepackage{bm}
\usepackage{bbm}
\usepackage{times}
\usepackage{nicefrac}
\usepackage{amsmath}
\usepackage{amsfonts}
\usepackage{amssymb}
\usepackage{amsthm}

\newcolumntype{.}{D{x}{}{-1}}

\newcommand{\vare}{\varepsilon}

\newcommand{\bfp}{{\bm {p}}}

\newcommand{\lbr}{\langle}
\newcommand{\rbr}{\rangle}

\newcommand{\Za}{Z\alpha}

\begin{document}

\title{Nuclear recoil effect in the Lamb shift of light hydrogen-like atoms}

\author{V. A. Yerokhin} \affiliation{Center for Advanced Studies,
        Peter the Great St.~Petersburg Polytechnic University, Polytekhnicheskaya 29,
        St.~Petersburg 195251, Russia}

\author{V. M. Shabaev} \affiliation{Department of Physics, St.~Petersburg State University,
7/9 Universitetskaya nab., St.~Petersburg, 199034 Russia}

\begin{abstract}

We report high-precision calculations of the nuclear recoil effect to the Lamb shift of
hydrogen-like atoms to the first order in the electron-nucleus mass ratio and to all orders in
the nuclear binding strength parameter $\Za$. The results are in excellent agreement with the
known terms of the $\Za$ expansion and allow an accurate identification of the nonperturbative
higher-order remainder. For hydrogen, the higher-order remainder was found to be much larger than
anticipated. This result resolves the long-standing disagreement between the numerical all-order
and the analytical $\Za$-expansion approaches to the recoil effect and completely removes the
second-largest theoretical uncertainty in the hydrogen Lamb shift of the $1S$ and $2S$ states.

\end{abstract}

\pacs{06.20.Jr, 31.30.jf, 21.10.Ft}

\maketitle

The hydrogen atom is one of the simplest and the most accurately measured atomic systems in
physics. The highest precision has been achieved for the $1S$-$2S$ transition, whose frequency was
measured \cite{parthey:11} with the relative uncertainty of $4.2\times 10^{-15}$ or $10$~Hz. In
future, it should be possible to increase the experimental accuracy even further, closely
approaching the $1.3$~Hz natural linewidth of the $2S$ level. A combination of measured transition
frequencies in hydrogen with sophisticated theoretical calculations provides the method for
determination of the Rydberg constant \cite{mohr:12:codata}, which is currently the most accurately
known fundamental constant. A co-product of the determination procedure of the Rydberg constant is
the proton charge radius.

The proton charge radius has recently received much attention, after the experiments on the muonic
hydrogen \cite{pohl:10,antognini:13} reported a large (7$\sigma$) unexplained difference of the
proton charge radius values extracted from the muonic hydrogen and the usual (electronic) hydrogen.
One of the possible explanation of this puzzle might be a yet undiscovered problem in the theory of
the electronic hydrogen. For this reason, investigations of possible inconsistencies in the
hydrogen theory are of particular importance today. In the present Letter we report calculations
that remove the second-largest theoretical uncertainty in the hydrogen theory and resolve a
long-standing discrepancy between the analytical and the numerical approach to the nuclear recoil
effect.

General expressions for the nuclear recoil effect to first order in the electron-nucleus mass ratio
$m/M$ and to all orders in the nuclear binding strength parameter $Z\alpha$ ($Z$ is the nuclear
charge number and $\alpha$ is the fine-structure constant) were obtained by one of us
\cite{shabaev:85,shabaev:88} (see also \cite{shabaev:98:rectheo}) and later rederived by other
authors \cite{yelkhovsky:94:xxx,pachucki:95,adkins:07}. Numerical calculations to all orders in
$\Za$ were reported in Refs.~\cite{artemyev:95:pra,artemyev:95:jpb,shabaev:98:jpb}. The results of
these calculations agreed well with the first terms of the $\Za$ expansion
\cite{salpeter:52,pachucki:95,golosov:95}. However, a significant disagreement was later found for
the higher-order $\Za$ expansion terms. Specifically, the $\Za$-expansion result for the
$(\Za)^7m/M$ contribution for the $1s$ state obtained within the leading logarithmic approximation
\cite{pachucki:99:prab,melnikov:99} yielded $-(18/\pi)\,(\Za)^7m^2/M$, whereas the all-order
numerical calculations \cite{shabaev:98:jpb} provided the result of comparable magnitude but of the
opposite sign, $(10/\pi)\,(\Za)^7m^2/M$. The disagreement, repeatedly mentioned in reviews of
hydrogen theory, notably, in Refs.~\cite{mohr:05:rmp,mohr:08:rmp,mohr:12:codata}, remained
unexplained for fifteen years. The 0.7~kHz difference between the all-order and the $\Za$-expansion
results for the hydrogen ground state is the second-largest theoretical error in the hydrogen Lamb
shift \cite{mohr:12:codata}, after the 2.0~kHz error due to the electron two-loop self energy.

The goal of the present Letter is to perform a non-perturbative (in $\Za$) calculation of the
nuclear recoil effect to the Lamb shift of $n=1$ and $n = 2$ energy levels in light hydrogen-like
atoms. In order to compare our results with the results of the $\Za$-expansion calculations, we aim
for a very high numerical accuracy and extend our calculations for fractional $Z$ as low as $0.3$.
The nonperturbative results for different $Z$'s are then fitted to the known form of the $\Za$
expansion, and the resulting expansion coefficients are compared with the analytical results. As a
result, we resolve the disagreement between the $\Za$-expansion and the all-order calculations of
the nuclear recoil effect. We demonstrate that the discrepancy was caused by unusually large
higher-order terms omitted in the $\Za$-expansion calculations. Specifically, the coefficient at
the single logarithm $\ln (\Za)^{-2}$ in order $(\Za)^7m/M$ is found to be 16 times larger than the
coefficient at the squared logarithm reported in Refs.~\cite{pachucki:99:prab,melnikov:99}. As a
result, the inclusion of the single logarithmic contribution changes drastically the
$\Za$-expansion result for the higher-order recoil effect.

The nuclear recoil effect to the Lamb shift of hydrogen-like atoms, to first order in $m/M$ and to
all orders in $Z\alpha$, is represented as a sum of four terms,
\begin{align}\label{eq:01}
\Delta E_{\rm rec} = \Delta E_{\rm L} + \Delta E_{\rm C} + \Delta E_{\rm tr(1)} + \Delta E_{\rm tr(2)}\,,
\end{align}
where $\Delta E_{\rm L}$ (the {\em low-order part}) is the recoil correction as can be derived from
the Breit equation , $\Delta E_{\rm C}$ (the {\em Coulomb part}) is the QED recoil correction
induced by the exchange of an arbitrary number of virtual Coulomb photons between the electron and
the nucleus, $\Delta E_{\rm tr(1)}$ and $\Delta E_{\rm tr(2)}$ (the {\em one-transverse-photon} and
{\em two-transverse-photons} parts, respectively) are the QED recoil corrections induced by the
exchange of one or two transverse photons and an arbitrary number of virtual Coulomb photons.

\begin{table*}
\caption{Nuclear recoil correction for the point nucleus, expressed in terms of
$P(\Za)$ and $G_{\rm rec}(\Za)$, $1/\alpha = 137.0359895$.
\label{tab:pnt}
}
\begin{ruledtabular}
  \begin{tabular}{l........}
$Z$ & \multicolumn{2}{c}{$1s$}
& \multicolumn{2}{c}{$2s$}
& \multicolumn{2}{c}{$2p_{1/2}$}
& \multicolumn{2}{c}{$2p_{3/2}$}
 \\
& \multicolumn{1}{c}{$P(\Za)$}
& \multicolumn{1}{c}{$G_{\rm rec}(\Za)$}
& \multicolumn{1}{c}{$P(\Za)$}
& \multicolumn{1}{c}{$G_{\rm rec}(\Za)$}
& \multicolumn{1}{c}{$P(\Za)$}
& \multicolumn{1}{c}{$G_{\rm rec}(\Za)$}
& \multicolumn{1}{c}{$P(\Za)$}
& \multicolumn{1}{c}{$G_{\rm rec}(\Za)$}
 \\
 \hline\\[-3pt]
  1    & 5.429\,x9035\,(2) & 9.72x0\,(3) &  6.155\,1x155\,(2) & 14.89x9\,(3) & -0.301\,1x22\,17\,(1) & 1.509\,x7\,(2)  & -0.301\,x316\,16\,(1) & -2.133x3\,(2)  \\
       & 5.429\,x90\,(3)^a & 9.7\,(x6)^a   &  6.154\,8x3\,(5)^a& 9.5\,(x9)^a & -0.301\,1x2^a         & 1.55\,(x9)^a    & -0.301\,x3\,(4)^b     & -2.\,(8.x)^b\\
       & 5.428\,x44\,^c    & -17.75x\,^c &  6.153\,3x8\,^c &-17.75x\,^c   &  -0.301\,2x0\,^c      &                 & -0.301\,x20\,^c  \\
  2    & 4.952\,x8246\,(3) & 10.39x0\,(1) & 5.678\,7x451\,(3) & 15.01x0\,(1) & -0.293\,2x82\,31\,(1) & 1.307\,x39\,(5) & -0.293\,x938\,24\,(1) & -1.772x02\,(5)   \\
  3    & 4.668\,x6482\,(5) & 10.48x03\,(9) & 5.395\,6x454\,(5) & 14.78x06\,(9)& -0.285\,3x47\,72\,(1) & 1.192\,x04\,(2) & -0.286\,x671\,66\,(1) & -1.570x41\,(2)\\
  4    & 4.464\,x0355\,(5) & 10.41x55\,(6) & 5.192\,4x455\,(5) & 14.49x26\,(6)& -0.277\,3x29\,22\,(2) & 1.112\,x68\,(2) & -0.279\,x498\,03\,(1) & -1.432x80\,(1)\\
  5    & 4.303\,x4275\,(5) & 10.29x44\,(4) & 5.033\,5x649\,(5) & 14.20x13\,(4)& -0.269\,2x33\,36\,(2) & 1.053\,x21\,(2) & -0.272\,x405\,66\,(2) & -1.329x68\,(2)
\end{tabular}
\end{ruledtabular}
$^a\,$ Shabaev et al. 1998 \cite{shabaev:98:jpb}\,,\ \ $^b\,$ Artemyev et al. 1995 \cite{artemyev:95:jpb}\,,\ \
$^c\,$ $\Za$ expansion.
\end{table*}

\textit{Point nucleus.} -- We first consider the nucleus to be the point source of Coulomb field.
In this case, the low-order part $\Delta E_{\rm L}$ is given by \cite{shabaev:85}
\begin{align} \label{eq:02}
\Delta E_{\rm L} =& \frac{1}{2M} \langle a \bigl| \bfp^2
 % \nonumber \\ &
- \boldsymbol{D}(0) \cdot \boldsymbol{p} - \boldsymbol{p} \cdot \boldsymbol{D}(0) | a
\rangle\,,
\end{align}
where $\bfp$ is the electron momentum operator, $D_j (\omega) = -4\pi \alpha Z \alpha_i D_{ij}
(\omega, r)$, and $D_{ij} (\omega, r)$ is the transverse part of the photon propagator in the
Coulomb gauge \cite{shabaev:98:rectheo}. Eq.~(\ref{eq:02}) can be calculated analytically in a very
simple form \cite{shabaev:85},
\begin{align} \label{eq:02a}
\Delta E_{\rm L} =&\ \frac{m^2 - \vare_a^2}{2M}\,,
\end{align}
where $\vare_a$ is the Dirac energy of the reference state.

The corrections $\Delta E_{\rm C}$, $\Delta E_{\rm tr,1}$, and $\Delta E_{\rm tr(2)}$ in
Eq.~(\ref{eq:01}) can be calculated only within the QED theory \cite{shabaev:85,
shabaev:88,yelkhovsky:94:xxx,pachucki:95, shabaev:98:rectheo,adkins:07}. The results for them are
\begin{align} \label{eq:10}
\Delta E_{\rm C} =
&\
\frac{2\pi i}{M}
\int_{-\infty}^\infty \mathrm{d}\omega\,
    \delta_{+}^2 (\omega) \,
%\nonumber \\ & \times
    \langle a | [\boldsymbol{p}, V] \, G(\omega + \vare_a)\,  [\boldsymbol{p}, V] | a \rangle\,,
 \\ \label{eq:11}
\Delta E_{\rm tr(1)} = &\
 - \frac{1}{M} \int_{-\infty}^\infty \mathrm{d}\omega\, \delta_{+} (\omega)\,
% \nonumber\\ & \times
    \langle a | \bigl\{ [\boldsymbol{p}, V] \,G(\omega + \vare_a)\boldsymbol{D}(\omega)\,
 \nonumber\\ &
    - \boldsymbol{D}(\omega)\, G(\omega + \vare_a) \, [\boldsymbol{p}, V] \big\} | a \rangle\,,
 \\ \label{eq:12}
\Delta E_{\rm tr(2)} = &\  \frac{i}{2\pi M} \int_{-\infty}^\infty
\mathrm{d}\omega \, \langle a | \boldsymbol{D}(\omega) \, G(\omega + E_a) \, \boldsymbol{D}(\omega) | a \rangle\,,
\end{align}
where the scalar product is implicit, $\delta_{+} (\omega) = i/(2\pi)/(\omega+i0)$, $V(r) = -\Za/r$
is the nuclear Coulomb potential, $G(\omega)$ is the relativistic Coulomb Green function, and $[.\
,\!\ .]$ denotes commutator.

For low-$Z$ ions, the QED part of the recoil effect can be conveniently parameterized as
\begin{align}\label{rec:1}
\Delta E_{\rm C} + \Delta E_{\rm tr(1)} + \Delta E_{\rm tr(2)} = \frac{m^2}{M}\, \frac{(\Za)^5}{\pi\, n^3}\,P(\Za)\,,
\end{align}
where $P(\Za)$ is a slowly-varying dimensionless functions, whose $\Za$ expansion is
\begin{align}
P(\Za) = &\ \ln (\Za)^{-2}\,D_{51} + D_{50}
  \nonumber \\ &
 + (\Za)\, D_{60}+ (\Za)^2\,G_{\rm rec}(\Za)\,,
\end{align}
and $G_{\rm rec}(\Za)$ is the higher-order remainder containing all higher orders in $\Za$,
\begin{align} \label{Zaho}
G_{\rm rec}(\Za) &\, = \ln^2 (\Za)^{-2}\,D_{72} + \ln (\Za)^{-2}\,D_{71} + D_{70} + \ldots\,.
\end{align}
The coefficients of the $\Za$ expansion are
\cite{salpeter:52,pachucki:95,golosov:95,pachucki:99:prab,melnikov:99}
\begin{align}
D_{51}  = &\ \frac13\,\delta_{l,0}\,,\ \  \ \ \
D_{50}  =  \biggl[ -\frac83\,\ln k_0(n,l) + d_{50}\biggr] \,,\\
D_{60}  = &\ \bigl( 4\ln 2-\frac{7}{2}\bigr)\pi\,\delta_{l,0}
%  \nonumber \\ &
+ \left[ 3- \frac{l(l+1)}{n^2}\right]\, \frac{2\pi(1-\delta_{l,0})}{(4l^2-1)(2l+3)}\,, \\
D_{72}  = &\  -\frac{11}{60}\,\delta_{l,0}\,,
\end{align}
where $\ln k_0(n,l)$ is the Bethe logarithm, whose numerical values are \cite{drake:90} $\ln
k_0(1s) = 2.984\,128\,556\,,$ $\ln k_0(2s) = 2.811\,769\,893\,,$ $\ln k_0(2p) =-0.030\,016\,709\,.$
The values of the coefficients $d_{50}$ for the states of interest are
\begin{align}
d_{50}(1s) &\ = \frac{14}{3}\ln 2 + \frac{62}{9}\,,\
d_{50}(2s)  = \frac{187}{18}\,,\
d_{50}(2p)  = -\frac{7}{18}\,.
\end{align}

The first numerical calculations of the QED recoil corrections (\ref{eq:10})-(\ref{eq:12}) to all
orders in $\Za$ were performed in Refs.~\cite{artemyev:95:pra,artemyev:95:jpb}. Results with an
improved precision were later reported for hydrogen in Ref.~\cite{shabaev:98:jpb}. The numerical
accuracy of these calculations, however, was not high enough. The general opinion
\cite{mohr:05:rmp,mohr:08:rmp,mohr:12:codata} was that these results were not fully consistent with
the higher-order $\Za$-expansion terms derived in Refs.~\cite{pachucki:99:prab,melnikov:99}. In the
present calculations we improve the numerical accuracy of the QED recoil corrections by 2-3 orders
of magnitude as compared to the previous studies.

The general scheme of the calculation was described in detail in Ref.~\cite{artemyev:95:pra}. The
Dirac-Coulomb Green function in Eqs.~(\ref{eq:10})-(\ref{eq:12}) was evaluated by summing over the
whole spectrum of the Dirac equation with the help of the finite basis set constructed with
$B$-splines \cite{johnson:88}. In the previous calculations \cite{artemyev:95:pra,shabaev:98:jpb},
the convergence of the results in the low-$Z$ region with increase of the basis size was hampered
by numerical instabilities associated with the limitations of the double-precision arithmetics. In
the present work we implemented the procedure of solving the Dirac equation with the $B$-splines
basis set in the quadruple-precision (32-digit) arithmetics. After that we were able to achieve a
clear convergence pattern of the calculated results when the size of the basis set was increased.
The largest basis size used in actual calculations was $N = 250$. The numerical uncertainty of the
obtained results was estimated by changing the size of the basis set by 30-50\% and by increasing
the number of integration points in numerical quadratures.

The numerical results for the $n= 1$ and $n= 2$ states and $Z = 1$-$5$ are presented in
Table~\ref{tab:pnt}. For the $1s$ and $2p_j$ states of hydrogen, we find perfect agreement with the
previous calculations \cite{artemyev:95:jpb,shabaev:98:jpb}. For the $2s$ state, there is a small
deviation caused by a minor mistake in the previous calculation. At the same time, we observe a
strong contrast between the all-order results for $G_{\rm rec}(1s)$ and $G_{\rm rec}(2s)$ and the
corresponding $\Za$-expansion results. We recall that the $\Za$-expansion results for $G_{\rm rec}$
include only the double-log contribution $D_{72} \ln^2 (\Za)^{-2}$ and neglect the higher-order
terms. For hydrogen, $\ln (1\alpha)^{-2} \approx 10$ is a large parameter, and the leading
logarithmic approximation is routinely used for estimating the tail of the $\Za$ expansion, with
typical uncertainty of 50\% \cite{mohr:05:rmp}.

\begin{figure}[t]
\centerline{
\resizebox{0.5\textwidth}{!}{%
  \includegraphics{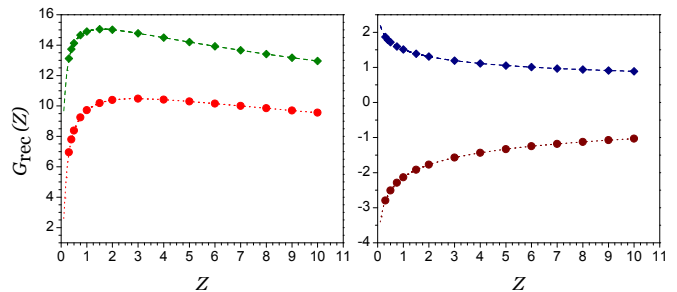}
}}
 \caption{(Color online)
Higher-order recoil correction $G_{\rm rec}(Z)$ for the $1s$ state (left, dots, red), the $2s$ state (left, diamonds, green),
the $2p_{1/2}$ state (right, diamonds, blue), and the $2p_{3/2}$ state (right, dots, brown).
 \label{fig:Grec}}
\end{figure}

In order to make a detailed comparison with the $\Za$-expansion results, we performed our
calculations for a series of $Z$ including fractional values as low as $Z = 0.3$. The results
obtained for the higher-order remainder $G_{\rm rec}(\Za)$ are plotted in Fig.~\ref{fig:Grec}. We
discover a rapidly changing structure at very low values of $Z$. Most remarkably, the bending of
the curve is practically undetectable for results with $Z \ge 2$. In order to access such a
structure in an all-order calculation, one needs to achieve a very high numerical accuracy at very
low (and fractional) values of $Z$.

Fitting our numerical results for the function $G_{\rm rec}(\Za)$ to the form of Eq.~(\ref{Zaho}),
we obtain results for the expansion coefficients. For the double-log contribution we find
$D_{72}(1s)= -0.183\,(1)$ and $D_{72}(2s)= -0.183\,(1)$, in agreement with the analytical value
$-11/60 \approx -0.1833$. The fitted results for the next two coefficients are:
\begin{align}
&\ D_{71}(1s)  = 2.919\,(10)\,,\ \ \
D_{70}(1s)  = -1.32\,(10)\,,\\
&\ D_{71}(2s)  = 3.335\,(10)\,,\ \ \
D_{70}(2s)  = -0.26\,(6)\,,\\
&\  D_{71}(2p_{1/2}) = 0.149\,(5)\,,\ \ \
D_{70}(2p_{1/2})  = -0.04\,(2)\,,\\
&\ D_{71}(2p_{3/2}) =  -0.283\,(5)\,,\ \ \
D_{70}(2p_{3/2})  = 0.69\,(2)\,.
\end{align}
We thus conclude that our all-order results are perfectly consistent with all known coefficients of
the $\Za$ expansion. The deviation observed for the $S$ states in Table~\ref{tab:pnt} comes from
the higher-order terms, whose contribution turns out to be unexpectedly large. Specifically, the
single-log coefficient $D_{71}$ is found to be 16 times larger than the double-log coefficient
$D_{72}$. As a result, the inclusion of the single-log contribution changes drastically the $\Za$
expansion result for the higher-order recoil effect.

%%%%%%%%%%%%%%%%%%%%%%%%%%%%%%%%%%%%%%%%%%%%%%%%%%%%%%%%%%%%%%%%%%%%%%%%
%%%%%
%%%%%
%%%%%%%%%%%%%%%%%%%%%%%%%%%%%%%%%%%%%%%%%%%%%%%%%%%%%%%%%%%%%%%%%%%%%%%

\begin{table*}
\caption{Finite nuclear size recoil correction, expressed in terms of
$\delta_{\rm fns} P$.
\label{tab:fns}
}
\begin{ruledtabular}
  \begin{tabular}{lc....}
$Z$ & $R$ [fm]
& \multicolumn{1}{c}{$1s$}
& \multicolumn{1}{c}{$2s$}
& \multicolumn{1}{c}{$2p_{1/2}$}
& \multicolumn{1}{c}{$2p_{3/2}$}
 \\
 \hline\\[-4pt]
  1    & 0.8775 & -0.000\,x1840\,(8)(9) & -0.000\,x1840\,(8)(8) & -0.000\,x000\,01         & -0.000\,x000\,01 \\
       &        &  0.000\,x0\,(2)^a \\
%       & 2.1424 & -0.000\,x786\,(1)(6)  & -0.000\,x787\,(1)(6)  & -0.000\,x000\,03\,(5)(0) & -0.000\,x000\,04 \\
  2    & 1.6755 & -0.000\,x628\,(4)(4)  & -0.000\,x629\,(4)(4)  & -0.000\,x000\,06\,(6)(0) & -0.000\,x000\,04\,(4)(0) \\
  3    & 2.4440 & -0.001\,x28\,(1)(1)   & -0.001\,x29\,(1)(1)   & -0.000\,x0002\,(4)(0)  & -0.000\,x000\,1\,(1)(0) \\
  4    & 2.5190 & -0.001\,x50\,(2)(1)   & -0.001\,x50\,(2)(1)   & -0.000\,x0003\,(7)(0)  & -0.000\,x000\,2\,(2)(0) \\
  5    & 2.4060 & -0.001\,x56\,(2)(1)   & -0.001\,x56\,(2)(1)   & -0.000\,x0004\,(10)(0)  & -0.000\,x0002\,(2)(0) \\
\end{tabular}
\end{ruledtabular}
$^a\,$ Shabaev et al. 1998 \cite{shabaev:98:recground}\,\\
\end{table*}

\textit{Finite nuclear size.} -- We now consider the correction to the nuclear recoil effect
induced by the finite nuclear size (fns), $ E_{\rm fns, rec} = E_{\rm rec}({\rm ext}) - E_{\rm
rec}({\rm pnt})\,, $ where $E_{\rm rec}({\rm ext})$ and $E_{\rm rec}({\rm pnt})$ are the recoil
corrections (\ref{eq:01}) evaluated with the extended and the point nuclear-charge distributions,
respectively. In the hydrogen theory \cite{mohr:12:codata}, the leading part of the recoil fns
effect is accounted for by introducing the reduced mass prefactor $[M/(m+M)]^3$ in the expression
for the fns correction. To the first order in $m/M$, the reduced-mass fns correction is
\begin{align} \label{eq:101}
E_{\rm fns, rm} = -3\,\frac{m}{M}[\vare_a({\rm ext})-\vare_a({\rm pnt})]\,,
\end{align}
where $\vare_a({\rm ext})$ and $\vare_a({\rm pnt})$ are the eigenvalues of the Dirac equation with
the extended and the point nuclear distributions, respectively. In the present Letter, we are
interested in the higher-order fns recoil correction $E^{\rm ho}_{\rm fns, rec}$ beyond the
reduced-mass part. It will be parameterized in terms of the function $\delta_{\rm fns}P$,
\begin{align}
E^{\rm ho}_{\rm fns, rec} \equiv E_{\rm fns, rec} - E_{\rm fns, rm} = \frac{m^2}{M}\, \frac{(\Za)^5}{\pi\, n^3}\,\delta_{\rm fns}P\,.
\end{align}
The fns recoil correction was studied in
Refs.~\cite{shabaev:98:recground,shabaev:98:ps,aleksandrov:15}. The numerical accuracy of these
studies, however,  was not sufficient for making any conclusions about the higher-order fns recoil
effect for hydrogen. In the present Letter, we perform the first high-precision evaluation of this
effect.

The low-order part of the recoil correction for an extended nuclear charge is given by
\cite{grotch:69,borie:82} (see also \cite{aleksandrov:15})
\begin{align} \label{eq:elexact}
\Delta E_{\rm L} = &\ \frac{1}{2M}\, \lbr a | \bigl[\vare_{a}^2 - m^2
\nonumber \\ &
    - 2m \beta V(r) - W^\prime (r) V^\prime(r) - V^2(r) \bigr] | a \rbr\,,
\end{align}
where $V^{\prime}(r) = {\rm d}V(r)/{\rm d}r$, $W^{\prime}(r) = {\rm d}W(r)/{\rm d}r$,
\begin{eqnarray}
V(r) &=& -\Za\int \mathrm{d}\boldsymbol{r}^\prime \frac{\rho(\boldsymbol{r}^\prime)}{|\boldsymbol{r} - \boldsymbol{r}^\prime|}, \label{eq:V}\\
W(r) &=& -\Za\int \mathrm{d}\boldsymbol{r}^\prime \rho(\boldsymbol{r}^\prime)|\boldsymbol{r} -
\boldsymbol{r}^\prime|, \label{eq:W}
\end{eqnarray}
 and $\rho
(\boldsymbol{r})$ is the density of the nuclear charge distribution $\big ( \int
\mathrm{d}\boldsymbol{r} \rho (\boldsymbol{r}) = 1\big )$. The Coulomb part of the recoil
correction $\Delta E_{\rm C}$ is given by the same formula (\ref{eq:10}) with $V(r)$ being the
potential of the extended nucleus (\ref{eq:V}). Exact expressions for the one-transverse-photon
part $\Delta E_{\rm tr(1)}$ and the two-transverse-photon part $\Delta E_{\rm tr(2)}$ for the
extended nucleus case are not yet known. In the present work, we will use the expressions
(\ref{eq:11}) and (\ref{eq:12}) derived for the point nucleus and evaluate the matrix elements with
the extended-nucleus potential $V(r)$, wave functions, energies, and propagators.

In order to estimate the uncertainty introduced by this approximation, we compare the low-order
part as evaluated in two ways: first, by the exact formula (\ref{eq:elexact}), $\Delta E_{\rm L}$,
and, second, using the operators derived for a point nucleus (see Eq.~(4) of
Ref.~\cite{shabaev:98:recground}), $\Delta E_{\rm L}^{\rm appr}$. We then estimate the
approximation error as the absolute value of
\begin{align} \label{eq:201}
2\,\frac{\Delta E_{\rm L}-\Delta E_{\rm L}^{\rm appr}}{E_{\rm fns, rec}^{\rm ho}}\left[
\Delta E_{\rm tr(1), fns} + \Delta E_{\rm tr(2), fns} \right]\,,
\end{align}
where  $\Delta E_{\rm tr(1), fns}$ and $\Delta E_{\rm tr(2), fns}$ are the fns corrections to the
one-transverse-photon and the two-transverse-photon parts, respectively. We note that one should
not use $\Delta E_{\rm L}$ in the denominator of Eq.~(\ref{eq:201}) because of a cancellation of
spurious terms between $\Delta E_{\rm L}$ and $\Delta E_{\rm C}$ \cite{shabaev:98:recground}. It
might be also mentioned that the full two-transverse-photon fns correction contains terms induced
by virtual nuclear excitations \cite{shabaev:98:rectheo,salpeter:52}. These terms should be
considered together with the nuclear polarization effect
\cite{khriplovitch:00,pachucki:07:heliumnp} and are beyond the scope of this Letter.

The numerical results for the higher-order fns recoil effect are presented in Table~\ref{tab:fns}.
The calculations for an extended nucleus were performed by the Dual kinetic balance method
\cite{shabaev:04:DKB} implemented in the quadruple-precision arithmetics. The results listed in the
table have two uncertainties, the first one representing the estimated uncertainty of the
approximation and the second one reflecting the dependence of the results on the nuclear model. In
order to estimate the model dependence, we performed our calculations for two models of the nuclear
charge distribution, the Gauss model \cite{yerokhin:15:Hlike} and the homogeneously charged sphere
model, with the same root-mean-square radii $R$. The results obtained with the Gauss model are
listed in the table, whereas the second uncertainty represents the absolute value of the difference
of the results of the two models. The error due to uncertainties of the nuclear radii is not
included in the table. It should be accounted for separately, e.g., by a simple estimate
$(2\,\delta R/R)\,\delta_{\rm fns} P \,,$ where $\delta R$ is the uncertainty of the radius $R$.

In summary, we calculated the nuclear recoil effect to the Lamb shift. The calculation is performed
to the first order in the electron-nucleus mass ratio $m/M$ and to all orders in the nuclear
binding strength parameter $\Za$, with inclusion of the finite nuclear size effect. The results
were shown to be in excellent agreement with the known terms of the $\Za$ expansion. The
higher-order recoil contribution beyond the previously known $\Za$-expansion terms was identified.

Our calculation resolves the previously reported disagreement between the numerical all-order and
the analytical $\Za$-expansion approaches and eliminates the second-largest theoretical uncertainty
in the hydrogen Lamb shift of the $1S$ and $2S$ states. For the point nucleus, the higher-order
correction beyond the previously known $\Za$-expansion terms for the hydrogen $1S$ and $2S$ Lamb
shift was found to be $0.65$ and $0.08$~kHz, respectively. In addition, we found the corresponding
shifts from the finite nuclear size recoil effect beyond the reduced mass of $-0.08$ and
$-0.01$~kHz. These results may be compared with the $0.01$~kHz experimental uncertainty of the
$1S$-$2S$ transition \cite{parthey:11}.

The higher-order recoil corrections are also important for the interpretation of experimental
results for the hydrogen-deuterium isotope shift. Indeed, the higher-order recoil corrections
calculated in the present Letter increase the theoretical value of the hydrogen-deuterium $1S$-$2S$
isotope shift by $0.36$~kHz (including $0.28$~kHz from the point nucleus and $0.08$~kHz from the
finite nuclear size), which may be compared with the experimental uncertainty of $0.015$~kHz
\cite{parthey:10} and the theoretical uncertainty of $0.6$~kHz \cite{jentschura:11}. The change of
the theoretical value increases the deuteron-hydrogen mean-square charge-radii difference as
obtained in Ref.~\cite{jentschura:11} by 0.00026~fm$^2$.

The results obtained in the present Letter demonstrate the importance of the non-perturbative (in
$\Za$) calculations as an alternative to the traditional $\Za$-expansion approach. Despite the
smallness of the parameter $\Za$ for hydrogen, $1\alpha \approx 0.0073$, the convergence of the
(semi-analytical) $\Za$ expansion is complicated by the presence of powers of logarithms. Moreover,
the predictive power of the $\Za$ expansion calculations is limited by the difficulty to reliably
estimate contributions of the uncalculated tail of the expansion.

V.A.Y. acknowledges support by the Russian Federation program for organizing and carrying out
scientific investigations. The work of V.M.S. was supported by RFBR (grant No. 13-02-00630) and by
SPbSU (grants No. 11.38.269.2014 and 11.38.237.2015).


\begin{thebibliography}{10}

\bibitem{parthey:11} C.~G. Parthey, A.~Matveev, J.~Alnis, B.~Bernhardt, A.~Beyer, R.~Holzwarth,
  A.~Maistrou, R.~Pohl, K.~Predehl, T.~Udem, T.~Wilken, N.~Kolachevsky,
  M.~Abgrall, D.~Rovera, C.~Salomon, P.~Laurent, and T.~W. H\"ansch,
\newblock Phys. Rev. Lett. {\bf 107}, 203001 (2011).

\bibitem{mohr:12:codata} P.~J. Mohr, B.~N. Taylor, and D.~B. Newell,
\newblock Rev. Mod. Phys. {\bf 84}, 1527 (2012).

\bibitem{pohl:10} R.~Pohl et~al.,
\newblock Nature (London) {\bf 466}, 213  (2010).

\bibitem{antognini:13} A.~Antognini et~al.,
\newblock Science {\bf 339}, 417  (2013).

\bibitem{shabaev:85} V.~M. Shabaev,
\newblock Teor. Mat. Fiz. {\bf 63}, 394  (1985)
\newblock [Theor. Math. Phys. {\bf 63}, 588 (1985)].

\bibitem{shabaev:88} V.~M. Shabaev,
\newblock Yad. Fiz. {\bf 47}, 107  (1988)
\newblock [Sov. J. Nucl. Phys. {\bf 47}, 69 (1988)].

\bibitem{shabaev:98:rectheo} V.~M. Shabaev,
\newblock Phys. Rev. A {\bf 57}, 59  (1998).

\bibitem{yelkhovsky:94:xxx} A.~Yelkhovsky,
\newblock arXiv: hep-th/9403095  (1994);
\newblock Report No. BudkerINP-94-27 (Budker Institute
  of Nuclear Physics, Novosibirsk, 1994).

\bibitem{pachucki:95} K.~Pachucki and H.~Grotch,
\newblock Phys. Rev. A {\bf 51}, 1854  (1995).

\bibitem{adkins:07} G.~S. Adkins, S.~Morrison, and J.~Sapirstein,
\newblock Phys. Rev. A {\bf 76}, 042508 (2007).

\bibitem{artemyev:95:pra} A.~N. Artemyev, V.~M. Shabaev, and V.~A. Yerokhin,
\newblock Phys. Rev. A {\bf 52}, 1884 (1995).

\bibitem{artemyev:95:jpb} A.~N. Artemyev, V.~M. Shabaev, and V.~A. Yerokhin,
\newblock J. Phys. B {\bf 28}, 5201 (1995).

\bibitem{shabaev:98:jpb} V.~M. Shabaev, A.~N. Artemyev, T.~Beier, and G.~Soff,
\newblock J. Phys. B {\bf 31}, L337  (1998).

\bibitem{salpeter:52} E.~E. Salpeter,
\newblock Phys. Rev. {\bf 87}, 328 (1952).

\bibitem{golosov:95} E.~A. Golosov, A.~S. Elkhovskii, A.~I. Milstein, and I.~B. Khriplovich,
\newblock Zh. Eksp. Teor. Fiz. {\bf 107}, 393 (1995)
\newblock [JETP {\bf 80}, 208 (1995)].

\bibitem{pachucki:99:prab} K.~Pachucki and S.~G. Karshenboim,
\newblock Phys. Rev. A {\bf 60}, 2792  (1999).

\bibitem{melnikov:99} K.~Melnikov and A.~S. Yelkhovsky,
\newblock Physics Letters B {\bf 458}, 143 (1999).


\bibitem{mohr:05:rmp} P.~J. Mohr and B.~N. Taylor,
\newblock Rev. Mod. Phys. {\bf 77}, 1  (2005).

\bibitem{mohr:08:rmp} P.~J. Mohr, B.~N. Taylor, and D.~B. Newell,
\newblock Rev. Mod. Phys. {\bf 80}, 633 (2008).

\bibitem{drake:90} G.~W.~F. Drake and R.~A. Swainson,
\newblock Phys. Rev. A {\bf 41}, 1243 (1990).

\bibitem{johnson:88} W.~R. Johnson, S.~A. Blundell, and J.~Sapirstein,
\newblock Phys. Rev. A {\bf 37}, 307  (1988).

\bibitem{shabaev:98:recground} V.~M. Shabaev, A.~N. Artemyev, T.~Beier, G.~Plunien, V.~A. Yerokhin,
    and
  G.~Soff,
\newblock Phys. Rev. A {\bf 57}, 4235  (1998).

\bibitem{shabaev:98:ps} V.~M. Shabaev, A.~N. Artemyev, T.~Beier, G.~Plunien, V.~A. Yerokhin, and
  G.~Soff,
\newblock Phys. Scr. {\bf T80}, 493  (1999).

\bibitem{aleksandrov:15} I.~A. Aleksandrov, A.~A. Shchepetnov, D.~A. Glazov, and V.~M. Shabaev,
\newblock J. Phys. B {\bf 48}, 144004 (2015).

\bibitem{grotch:69} H.~Grotch and D.~R. Yennie,
\newblock Rev. Mod. Phys. {\bf 41}, 350  (1969).

\bibitem{borie:82} E.~Borie and G.~A. Rinker,
\newblock Rev. Mod. Phys. {\bf 54}, 67  (1982).

\bibitem{khriplovitch:00} I.~Khriplovich and R.~A. Sen'kov,
\newblock Phys. Lett. B {\bf 481}, 447 (2000).

\bibitem{pachucki:07:heliumnp} K.~Pachucki and A.~M. Moro,
\newblock Phys. Rev. A {\bf 75}, 032521 (2007).

\bibitem{yerokhin:15:Hlike} V.~A. Yerokhin and V.~M. Shabaev,
\newblock J. Phys. Chem. Ref. Data {\bf 44}, 033103 (2015).

\bibitem{shabaev:04:DKB} V.~M.~Shabaev, I.~I.~Tupitsyn, V.~A.~Yerokhin, G.~Plunien, and G.~Soff,
\newblock Phys. Rev. Lett. {\bf 93}, 130405 (2004).

\bibitem{parthey:10} C.~G. Parthey, A.~Matveev, J.~Alnis, R.~Pohl, T.~Udem, U.~D. Jentschura,
  N.~Kolachevsky, and T.~W. H\"ansch,
\newblock Phys. Rev. Lett. {\bf 104}, 233001 (2010).

\bibitem{jentschura:11} U.~Jentschura, A.~Matveev, C.~Parthey, J.~Alnis, R.~Pohl, T.~Udem,
  N.~Kolachevsky, and T.~H\"ansch,
\newblock Phys. Rev. A {\bf 83}, 042505 (2011).

\end{thebibliography}
\end{document}